\documentclass{aa}
\usepackage{epsfig}
\begin{document}

\newcommand{\bea}{\begin{eqnarray}}    
\newcommand{\eea}{\end{eqnarray}}      
\newcommand{\be}{\begin{equation}}
\newcommand{\ee}{\end{equation}}
\newcommand{\hmp}{\,h^{-1}_{100}\,{\rm \mbox{Mpc}}}
\newcommand{\bef}{\begin{figue}}
\newcommand{\eef}{\end{figure}}
\newcommand{\etal}{et al.}
\newcommand{\kms}{\,{\rm km}\;{\rm s}^{-1}}
\newcommand{\hubunits}{\,\kms\;{\rm Mpc}^{-1}}
\newcommand{\hmpc}{\,h^{-1}\;{\rm Mpc}}
\newcommand{\hkpc}{\,h^{-1}\;{\rm kpc}}
\newcommand{\msun}{M_\odot}
\newcommand{\K}{\,{\rm K}}
\newcommand{\cm}{{\rm cm}}
\newcommand{\cd}{{\langle n(r) \rangle_p}}
\newcommand{\Mpc}{{\rm Mpc}}
\newcommand{\kpc}{{\rm kpc}}
\newcommand{\xir}{{\xi(r)}}
\newcommand{\xrp}{{\xi(r_p,\pi)}}
\newcommand{\xsirpi}{{\xi(r_p,\pi)}}
\newcommand{\wrp}{{w_p(r_p)}}
\newcommand{\gr}{{g-r}}
\newcommand{\Navg}{N_{\rm avg}}
\newcommand{\Mmin}{M_{\rm min}}
\newcommand{\fiso}{f_{\rm iso}}
\newcommand{\Mr}{M_r}
\newcommand{\rp}{r_p}
\newcommand{\zmax}{z_{\rm max}}
\newcommand{\zmin}{z_{\rm min}}

\def\eg{{e.g.}}
\def\ie{{i.e.}}
\def\spose#1{\hbox to 0pt{#1\hss}}
\def\ltapprox{\mathrel{\spose{\lower 3pt\hbox{$\mathchar"218$}}
\raise 2.0pt\hbox{$\mathchar"13C$}}}
\def\gtapprox{\mathrel{\spose{\lower 3pt\hbox{$\mathchar"218$}}
\raise 2.0pt\hbox{$\mathchar"13E$}}}
\def\inapprox{\mathrel{\spose{\lower 3pt\hbox{$\mathchar"218$}}
\raise 2.0pt\hbox{$\mathchar"232$}}}

\title{The zero-crossing scale of the galaxy correlation function
and the problem of galaxy bias}

\subtitle{}

\author{Francesco Sylos Labini \inst{1,2}}

\titlerunning{The zero-crossing scale and the problem of galaxy bias} 

\institute{ ``Enrico Fermi Center'', Via Panisperna 89 A, Compendio
del Viminale, 00184 Rome, Italy
\and``Istituto dei Sistemi Complessi'' CNR, Via dei Taurini 19, 00185
Rome, Italy}

\date{Received / Accepted}

\abstract{One of the main problems in the studies 
of large scale galaxy structures concerns the relation of the
correlation properties of a certain population of objects with those
of a selected subsample of it, when the selection is performed by
considering physical quantities like luminosity or mass. I consider
the case where the sampling is defined as in the simplest thresholding
selection scheme of the peaks of a Gaussian random field as well as
the case of the extraction of point distributions in high density
regions from gravitational N-body simulations.  I show that an
invariant scale under sampling is represented by the zero-crossing
scale of $\xi(r)$.  By considering recent measurements in the 2dF and
SDSS galaxy surveys I note that the zero-point crossing length has not
yet been clearly identified, while a dependence on the finite sample
size related to the integral constraint is manifest.  I show that this
implies that other length scales derived from $\xi(r)$ are also
affected by finite size effects. I discuss the theoretical
implications of these results, when considering the comparison of
structures formed in N-body simulations and observed in galaxy
samples, and different tests to study this problem.
\keywords{Cosmology: large-scale structure of Universe; 
Cosmology: dark matter Cosmology: observations }}
\maketitle

\section{Introduction}

The problem of ``sampling'' discrete and continuous distributions is a
central one in studies of cosmological density fields and particularly
of galaxy structures. By sampling I mean the operation performed when
one extracts, from a given distribution, a subsample of it by making a
selection on a certain parameter $\mu$. For example, one can make such
type of selection by extracting from the whole population of galaxies
of all luminosity, only those objects whose luminosity is brighter
than a given threshold. A similar selection can be done by considering
galaxy color. Alternatively one may consider a certain density field,
continuous or discrete, where the fluctuation field is a stochastic
variable of position (for example a Gaussian fluctuation field), and
one may sample the distribution by extracting fluctuations larger than
a given threshold in the density fluctuation.

In general the problem consists in the understanding the relations
between the statistical properties of the ``biased'' distribution with
the original one, particularly of the two-point correlation function
$\xi(r;\mu>\tilde \mu)$ (where $\tilde \mu$ is the threshold) of the
sampled field with the original $\xi(r;\mu)$. The interest, for
instance, lies in the fact that in the studies of galaxy samples, one
has to perform a sampling when measuring the two-point correlation
function. In the comparison of observation with theoretical models the
sampling procedure plays a crucial role in the determination of the
physics of the system. In fact, in the analysis of cosmological N-body
simulations one also needs to extract subsamples of points which,
according to some models, would represent galaxies instead of dark
matter particles. In these contexts, the simplest theoretical model
describing biasing (introduced by Kaiser 1984) is not able to take
into account the effects related to strong clustering, as it was
developed for a continuous Gaussian field, and thus it does not
represent an useful analytical treatment of the problemof strong
clustering, which is instead the relevant one for galaxy
structures. We show however that an important feature of this model is
preserved also in cases where strong clustering in point distributions
is present.

It is very difficult to treat the problem of sampling for a generic
case. What one can do realistically is to consider a certain point
distribution, with given correlation properties and a certain sampling
procedure and then look for invariant quantities under sampling, such
as characteristic length scales which are unaffected by sampling. This
is the strategy I am going to consider in this paper.

In this paper I firstly briefly review (Sec.2) the effect of sampling
in the simplest model of a correlated Gaussian density field. In Sec.3
I show that for the case of a Cold Dark Matter (CDM) type model such a
sampling does not change the intrinsic length scale defined by
$\xi(r_{zp};\mu)=0$, while other length scales are affected, in a
linear or non-linear way depending on scales and amplitudes.  I then
consider in Sec.4 particle distributions obtained from cosmological
N-body simulations extracted in such a way to represent large
amplitude fluctuations ultimately associated to galaxies in some
models. I show that also in this case the scale $r_{zp}$ remains
invariant under sampling, while, for example the scale such that
$\xi(r_0;\mu>\tilde \mu)=1$ changes as a function of the threshold
$\tilde \mu$. An important point related to finite sample measurements
of the correlation function is discussed in Sec.5: that is the problem
of the determination of the zero-point in relation to the estimators
of $\xi(r)$ and the finite-size effects which may artificially force
the correlation function to cross zero, even when the underlying
distribution, in the ensemble sense, has, for example, only positive
correlations: In this case the scale $r_{zp}$ is a finite size
effect. I consider in Sec. 6 the observational situation, also in the
light of the recent results of Eisenstein et al. (2005) on a very
large and deep sample of the Sloan Digital Sky Survey (SDSS).  I
discuss the fact that in different galaxy samples the length scale
$r_{zp}$ is not found to be stable, varying from 20 Mpc/h in the CfA1
catalog to about 120 Mpc/h in the SDSS data.  The conclusions are
discussed in Sec.7: I find that, contrary to the theoretical CDM case
and to results in N-body simulations, observational evidences support
the finite-size interpretation of the zero-point crossing scale of the
estimated $\xi(r;\mu>\tilde\mu)$. The case for such a variation can be
directly clarified by studying the conditional average density.  I
then discuss the implications concerning other length-scales measured
by the estimated correlation function, such as the scale where
$\xi(r_0;\mu>\tilde \mu)=1$, concluding that, in galaxy samples,
finite size effects may play the dominant role for their
determination. Finally I discuss some direct tests to clarify the
situation.


\section{Sampling a Gaussian random field} 

Let us now discuss the simplest biasing scheme of a continuous and
correlated density field, introduced by Kaiser (1984).  Suppose to
have a Gaussian random field with correlations described by
$\xi(r;\mu)$ and such that the variance is $\langle \mu^2 \rangle =
\sigma^2$ (where $\mu$ is the mean density normalized 
fluctuation). One can identify fluctuations of the field such that
they are larger than $\nu$ times the variance. This selection defines
a biased field with equal weight: 0 if the fluctuations of the
original field are smaller than $\tilde\mu\equiv \nu\sigma$ and 1 if
they are equal or larger than $\tilde\mu$. When one changes the
threshold $\nu$ one selects different regions of the underlying
Gaussian random field, corresponding to fluctuations of differing
amplitudes. The reduced two-point correlation function of the selected
objects is then that of the peaks $ \xi(r;\mu>\tilde\mu)$, which is
enhanced with respect to that of the underlying density field
$\xi(r;\mu)$ (normalized to $\sigma^2$). One may compute the following
first-order approximation (\cite{bias2})
\be
\xi(r;\mu>\tilde\mu)\approx 
\sqrt{\frac{1+\xi(r;\mu)}{1-\xi(r;\mu)}}\exp\left(\nu^2
\frac{\xi(r;\mu)}{1+\xi(r;\mu)}\right)-1
\label{approx-xinu} \;, 
\ee
which reduces to $\xi(r;\mu>\tilde\mu) \simeq \nu^2 \xi(r;\mu)$ when
$\nu^2 \xi(r;\mu)
\ll 1$.  Thus, if present in the underlying distribution,
 the characteristic length scale $r_{zp}$ is not changed under this
 selection procedure, i.e.
\be
\label{zpk}
\xi(r_{zp};\mu) = \xi(r_{zp};\mu>\tilde\mu)=0  \;\; \forall \tilde \mu \;.
\ee
On the other hand for $\xi(r;\mu>\tilde\mu) >1$ the amplification is
non-linear as a function of scale: this means that the functional
behavior of $\xi(r;\mu>\tilde\mu)$ is different from the one of
$\xi(r;\mu)$ in the regime where $\xi(r;\mu>\tilde\mu) >1$.  In
addition the scale such that $\xi(r_0;\mu>\tilde\mu )=1$ changes in a
non-linear way as a function of the threshold (\cite{bias1,bias2}).

\section{Sampling a CDM type density field} 

I discuss now the effect of the previous biasing scheme on a
cosmological relevant density field.  It has been discussed in
Gabrielli, Joyce \& Sylos Labini (2002) that main features of
correlated (Gaussian) density fields in standard cosmological models
can be captured by the following behavior of the power spectrum of
mean density normalized density fluctuations
$P(k;\mu)= Ak\exp(-k/k_c) \;,$
where $A$ is a constant and $k_c$ is the characteristic wave-number of
the ``turn-over'' scale. Its Fourier transform, the real space
two-point reduced correlation function, has the following behavior:
\be
\label{xi-cdmlike}
\xi(r;\mu)=\frac{A}{\pi^2} \frac{\left( \frac{3}{k_c^2} - r^2\right)}
{\left( \frac{1}{k_c^2}+r^2\right)^3}  \;,
\ee
where $\mu$ is now the value of the normalized density fluctuation.
One may consider the characteristic length scale $r_{zp}$, such that
$\xi(r_{zp};\mu)=0$, i.e.
\be
r_{zp} = \frac{\sqrt{3}}{k_c} \;.
\ee 
Other length scales can be defined to be dependent on the amplitude of
$\xi(r;\mu)$: For example one may identify the scale at which
$\xi(r;\mu)$ has a certain (positive) value (which in this context has
to be smaller than one by definition, as this is a continuous Gaussian
random density field) and thus identifying a length scale which will
be dependent on the amplitude $A$. According to Eq.\ref{zpk} the scale
$r_{zp}$ is invariant under the biasing scheme discussed in the
previous section (see Fig.\ref{xi.cdm.bias})

\begin{figure}
\begin{center}
\includegraphics*[angle=0, width=0.5\textwidth]{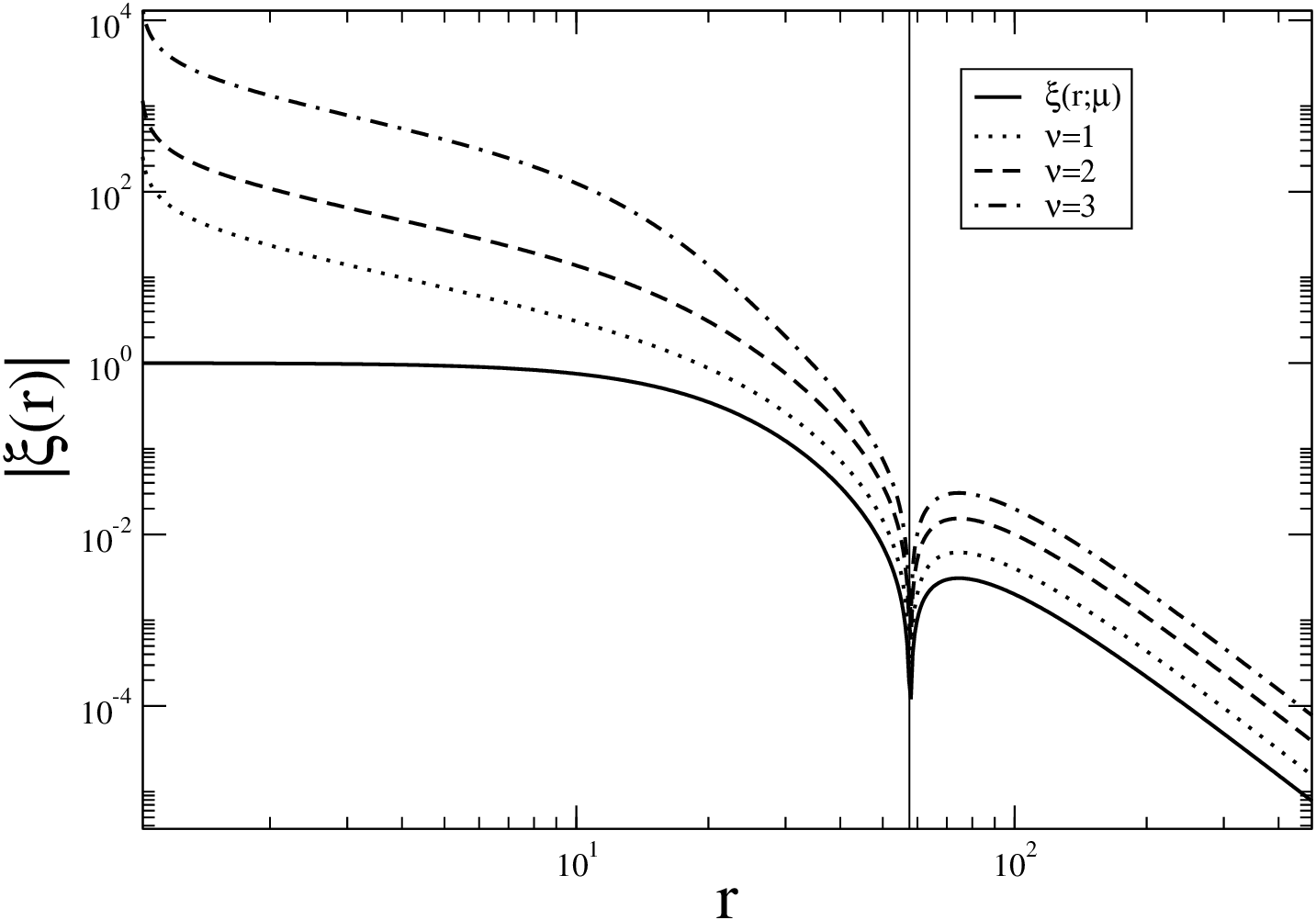}
\end{center}
\caption{Absolute value of the reduced 
correlation function of the toy model described by Eq.\ref{xi-cdmlike}
(solid line) and of the ones corresponding to different values of the
threshold parameter $\nu$ calculated by applying
Eq.\ref{approx-xinu}. The amplification is non-linear at small scales,
where $\xi(r;\mu>\tilde \mu) >1$, linear at large scales, and the
zero-crossing scale is invariant under biasing.}
\label{xi.cdm.bias}
\end{figure}

The correlation function given by Eq.\ref{xi-cdmlike} is different
from the one of a more realistic CDM model in the behavior at scales
for $r<r_{zp}$: In the CDM model in that range of scales $\xi(r;\mu)$
has an approximate power-law behavior of the type $\xi(r;\mu) \sim
r^{-1.5}$ with the introduction of some other  characteristic
scales.  However the zero-crossing scale $r_{zp}$ is still a clear
intrinsic feature which is not changed by the biasing scheme discussed in
the previous section.  At large scales $r > r_{zp}$ the CDM reduced
correlation function has the same $-r^{-4}$ behavior as
Eq.\ref{xi-cdmlike}. Both satisfy the important constraint
\be
\label{shcond}
\int_0^{\infty} \xi(r;\mu) r^2 dr =0 
\ee
which has been called ``the super-homogeneous condition'', in order to
make clear the fact that this corresponds to a global condition on the
correlation properties of particular systems which display a sort of
long-range order, or, alternatively, they are more ordered than purely
uncorrelated stochastic processes (e.g. Poisson) (\cite{glass}).
	

\section{Sampling points in cosmological N-body simulations}

Gravitational clustering in the regime of strong fluctuations is
usually studied through gravitational N-body simulations.  The
particles are not meant to describe galaxies but collision-less
dark-matter mass tracers (but see discussion in e.g.
\cite{grav1}). During gravitational evolution complex non-linear 
dynamics make non-linear structures at small scales, while at large
scales it occurs a linear amplification according to linear
perturbation theory. Thus, while on large scales correlation
properties do not change from the beginning --- a part a simple linear
scaling of amplitudes --- at small scales non-linear correlations are
built.  Typically in these simulations non-linear clustering is formed
up to scales of order of few Mpc (see e.g. \cite{bjsl}) and the
intrinsic scale $r_{zp}$ is unchanged, as typically $r_{zp} > 50$
Mpc/h in CDM models (see e.g. \cite{glass}).

At late times one can identify subsamples of points which trace the
high density regions, and these would  represent the
``galaxies'' whose statistical properties are ultimately compared with
the ones found in galaxy samples. Here I consider the GIF galaxy
catalog (\cite{gif}) constructed from a $\Lambda$CDM simulation run by
the Virgo consortium (\cite{virgo}).  The way in which this is done is
to firstly identify the halos, which represent almost spherical
structures with a power-law density profile from their center. The
number of galaxies belonging to each halo is set proportional to the
total number of points belonging to the halo to a certain power.
This procedure identifies points lying in high density regions of
the dark-matter particles.  One may assign to each point a luminosity
and a color on the basis of a certain criterion which is not relevant
for what follows (see
\cite{sheth} and reference therein). 
The resulting catalog is divided into two subsamples based on
``galaxy'' color as in Sheth et al. (2001): (brighter) red galaxies
(for which B-I is redder than 1.8) and (fainter) blue galaxies (B-I
bluer than 1.8).

In summary four samples of points may be considered: (i) the original
dark matter particles with N=$256^3$ particles (ii) all galaxies with
N=15445 (iii) blue galaxies with N=11023 and (iv) red galaxies with
N=4422. In Fig.\ref{xisimu} the behavior of $\xi(r)$ for the different
objects is shown
\footnote{The estimator of $\xi(r)$ is the full-shell one --- see 
Eq.\ref{xifs1} below}.
\begin{figure}
\begin{center}
\includegraphics*[angle=0, width=0.5\textwidth]{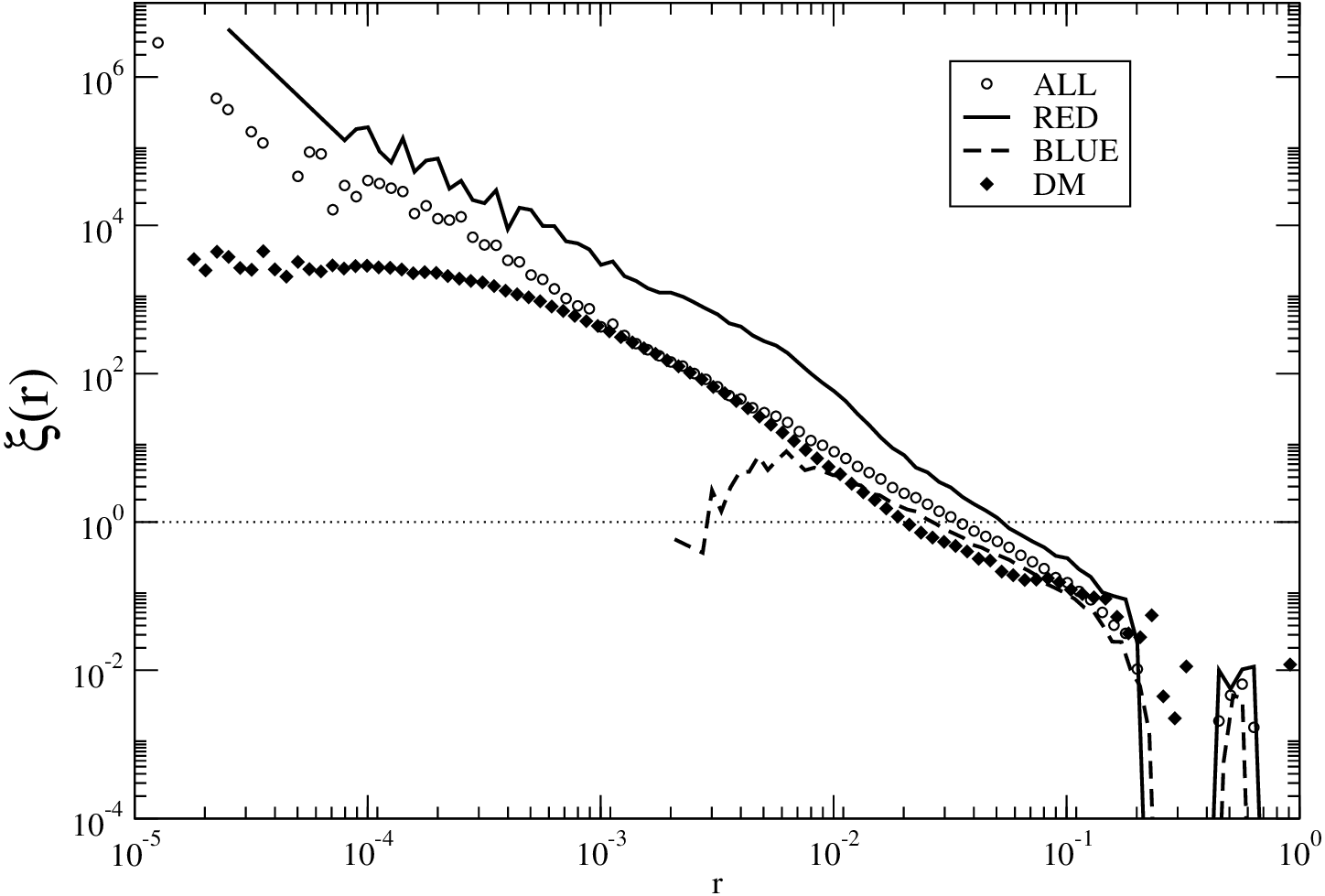}
\end{center}
\caption{Reduced correlation function for the four samples of 
points selected in the simulation: the original dark matter (DM)
field, all ``galaxies'' (ALL), blue galaxies (BLUE) and red
galaxies (RED). Distances are given in units of the sample size L=141.3
Mpc/h. }
\label{xisimu}
\end{figure}
One may notice that $\xi(r)$ for red (blue) galaxies has a 
larger (smaller) amplitude than the one of the original sample (all
galaxies). The underlying dark matter particles show almost the same
amplitude as all galaxies, although a change of slope at small scales
is manifest. The amplification is linear, i.e.  $\xi(r)$ for red
galaxies shows almost the same functional behavior of that of all
galaxies but with a larger amplitude.  Clearly the original point
distribution is not Gaussian, at least in the relevant range of scales
considered, but characterized by strong fluctuations and thus one
should explain such a mechanism of amplification (or de-amplification)
differently from what has been proposed by Kaiser (1984). On the other
hand the scale where the power-law behavior breaks down, and thus 
the scale $\xi(r_{zp})=0$, is invariant under sampling as for the
simple Gaussian threshold biasing scheme discussed above: the
amplitude {\it independent} characteristic scale is not changed under
biasing. The biasing mechanism described above does not introduce new
length scales in the system or change the intrinsic one, but it does
alter the amplitude of the average density and thus any scale
dependent on it (e.g. the scale such that $\xi(r_0)=1$).

Note that the zero-crossing scale of $\xi(r)$ cannot be in general
well established because of statistical fluctuations which affect any
finite sample estimation of correlations. In this case however a clear
signature of the zero-crossing scale is given by the sharp cut-off of
the reduced correlation function, in a log-log plot, at the scale of
order $r_{zp}$. This happens when the amplitude of the estimated
$\xi(r)$ is about $10^{-2}$, so that statistical noise does not affect
the measurement in a substantial way.  In the case considered, in
fact, the regime changes from being positively correlated, and larger
than unity, to small anti-correlation. This is the way used hereafter
to define the scale $r_{zp}$. In the general case, where the
functional behavior of the correlation function is more complicated
(e.g. with a very slow approach to zero) the way the zero-crossing
scale is estimated must be clearly explored.

In order to test the reality of the zero-crossing scale, one may cut
the sample at the scale $R_s\approx 0.3$ (in units normalized to the
box side) and recompute the correlation function.  No sensible change
is found in the scale $r_{zp}$. As discussed below, this happens
because the conditional density for scales $r>0.3$ is very well
approximated by a flat behavior corresponding to the transition from
strong to weak clustering, and the scale $r_{zp}$ is related, in this
case, to the scale where the conditional density flattens.

Note that in the regime where $\xi(r)\gg 1$ no clear a priori
prediction can be formulated on the amount of increase of amplitude of
$\xi(r)$ with sampling. Actually the perspective on this problem is to
choose a selection procedure such that it gives results similar to
what is found in galaxy catalogs. Thus the observations are used to
tune the selection in the simulations. The idea is in fact that one
may change the way points are selected up to when a satisfactory
agreement with what is observed in galaxy catalogs has been found.
This can be true for the strongly correlated regime, but the selection
employed does not change $r_{zp}$ which thus becomes the main length
scale to be studied when relating observed galaxy distributions  to
simulations and ultimately to the distribution of the underlying
dark matter particles.

In order to understand the origin of the amplification observed in the
sampled point distributions it is useful to study the behavior of the
conditional density which has a straightforward interpretation in
terms of correlations (see Fig.\ref{gammasimu}). This statistical tool
gives the average number of points observed in an infinitesimal shell
as a function of distance from a point of the distribution (and thus
this is a conditional quantity) and can be written as (e.g. Gabrielli
et al. 2004)
\be
\label{cd}
\cd  \equiv \frac{\langle n(\vec{r})n(0) \rangle}{\langle n(0) \rangle} \;, 
\ee 
where $n(\vec{r})$ is the microscopic particle number density. This is
related to $\xi(r)$ by the equation
\be 
\label{xi-gamma} \xi(r) \equiv
\frac{\cd}{n_0}-1 \;.  
\ee 
being $n_0>0$ the ensemble average density of the distribution.

\begin{figure}
\begin{center}
\includegraphics*[angle=0, width=0.5\textwidth]{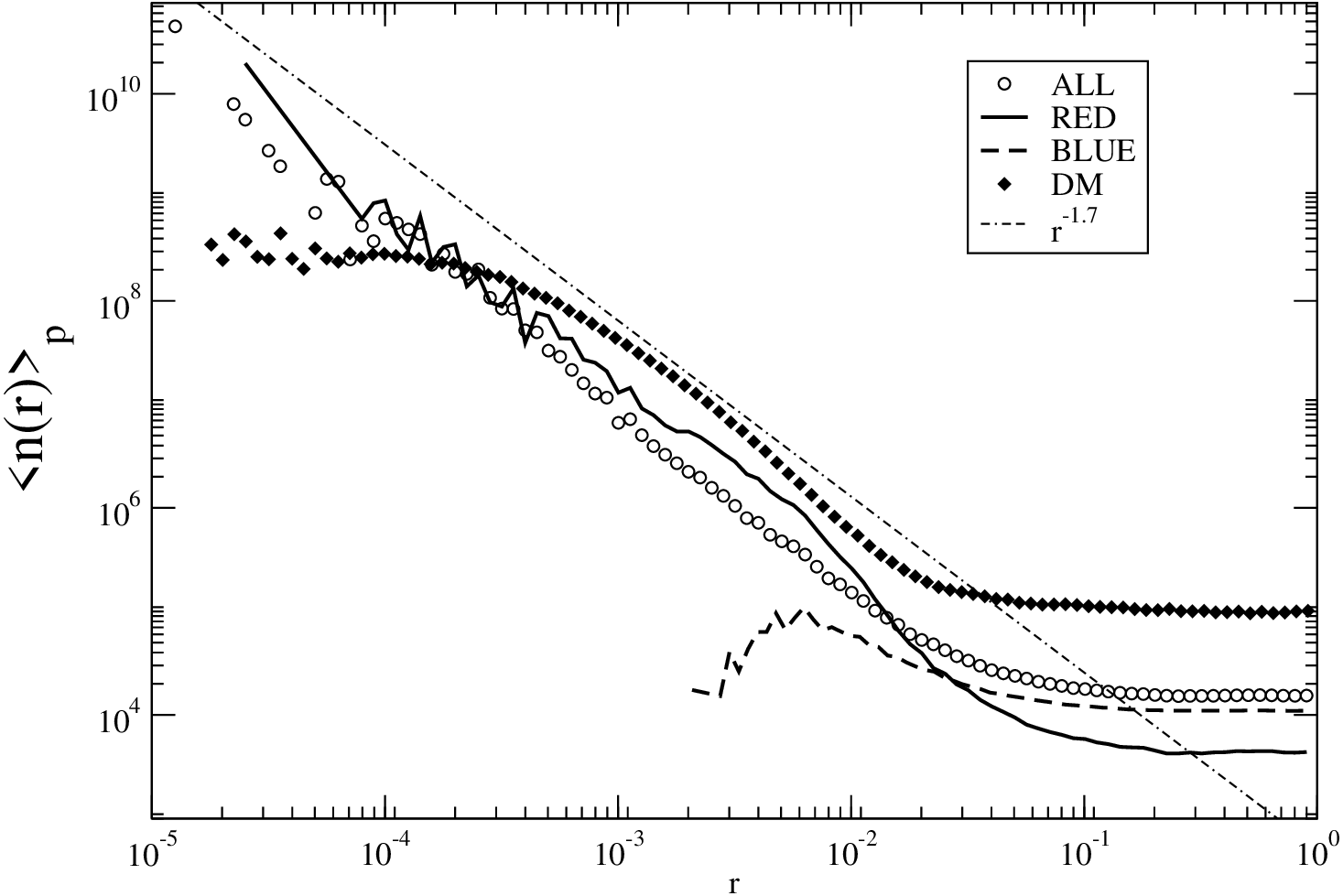}
\end{center}
\caption{Conditional density 
 for the samples of points shown in Fig.\ref{xisimu}. 
The conditional density for dark matter particles (DM) 
has been normalized arbitrarily. The reference line has a slope
$\gamma=1.7$}
\label{gammasimu}
\end{figure}
The red galaxies are responsible for the strong correlations observed
in the full sample as the conditional density is almost the same as
for all galaxies at small scales. At large scales there is instead a
fast decrease as the sample average of red galaxies is smaller than
the one of all galaxies (there are less objects). The amplification of
$\xi(r)$ of the red galaxies with respect to the full sample can be
explained as an almost constant value of the conditional density at
small scales together with a decrease of the sample density. It
follows from Eq.\ref{xi-gamma} that the amplitude of $\xi(r)$ is
amplified if $\cd$ remains the same and $n_0$ is lowered. This means
that for red galaxies the sampling is local, i.e. their conditional
density is (almost) invariant at small scales.  Clearly, as there are
globally less objects, the sample density of red galaxies is smaller
than that of all galaxies. On the other hand blue galaxies present only some 
residual correlations a small scales, and they are more numerous than
red galaxies.

The main conclusion is that the intrinsic characteristic length of the
model given by $r_{zp}$ (measured as discussed above) is not changed
by this selection procedure, in close analogy with what happens in the
simple Gaussian thresholding biasing scheme of a CDM field discussed
in the previous section.

\section{Finite size effects and the integral constraint}

Concerning the study of the zero-crossing scale of $\xi(r)$ a point 
must be clarified in relation to the estimator of this statistical
quantity.  Suppose that one chooses the so-called full-shell estimator 
(\cite{book})
defined as 
\be
\label{xifs1}
\xi_E(r) = \frac{\cd^E}{n_E} -1
\ee
where $n_E$ is the density in a sphere of radius $R_s$ up to which
$\cd^E$ can be estimated. 
As the sample density is estimated by 
\be
\label{ne1}
n_E= \frac{3}{4\pi R_s^3} \int_0^{R_s} \cd d^3r 
\ee
it follows that 
\be
\label{xifs}
\int_0^{R_s} \xi_E(r) r^2 dr = 0 \;.
\ee
This condition holds independently on $R_s$ and the true $\xi(r)$:
Thus in a finite sample one finds the zero crossing of $\xi_E(r)$ no
matter which are the true correlation properties of the
distribution\footnote{Note that Eq.\ref{xifs} holds only for the
full-shell estimator of $\xi(r)$. However, as discussed in
(\cite{glass,book}) similar boundary conditions, related to the fact
that the average density has been estimated inside a given sample,
must be verified by any estimator of $\xi(r)$.}. For example $\xi(r)$
can be a simple positive power-law extending to scales much larger
than $R_s$: its estimator in a finite sample will obey to
Eq.\ref{xifs}. The point to study is whether the zero-crossing scale
depends, or not, on the sample volume.

In the case of a CDM-like correlation function, where a
similar constraint holds in the whole space (see Eq.\ref{shcond}) one
can distinguish between the following behaviors (for simplicity, we
neglect in the following discussion the effect of statistical noise in
the estimator): (i) $R_s < r_{zp}$ --- in this case the positively
correlated range of scales at small scales will not be detected
entirely, but an artificial zero-point will be introduced at scales
comparable to $R_s$. In addition the amplitude of the estimator
$\xi_E(r)$ is scale dependent. (ii) $R_s > r_{zp}$ --- in this case the
zero crossing scale will be well-defined, in the sense that changing
$R_s$ the distance scale $r_{zp}$ will not change.  Hoverer the negative
correlated range of scales (i.e. $r > r_{zp}$) will be distorted (and the
absolute value of $\xi_E(r)$ is increased) by the condition
Eq.\ref{xifs} (see Fig.\ref{xicdm}).
\begin{figure}
\begin{center}
\includegraphics*[angle=0, width=0.5\textwidth]{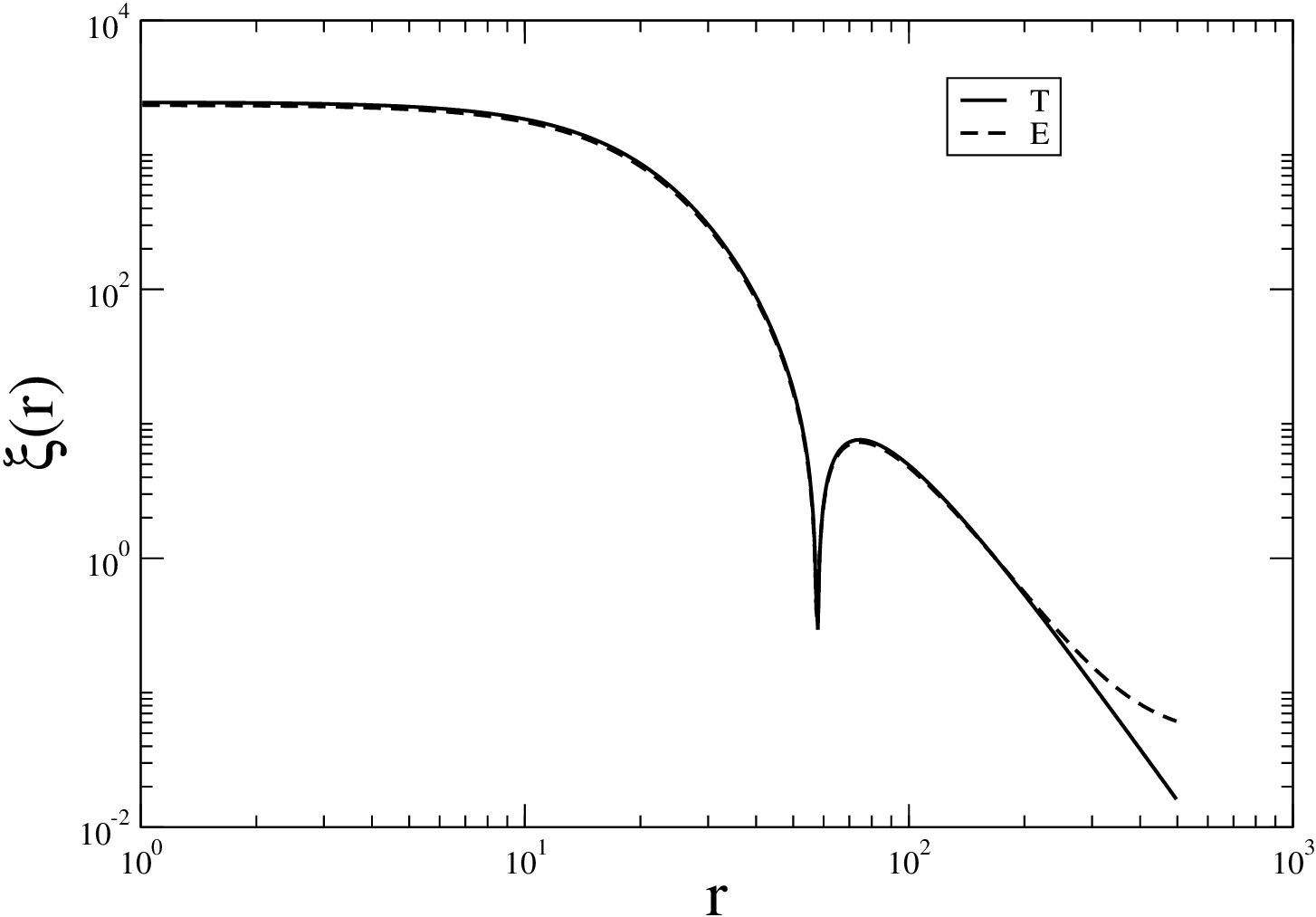}
\end{center}
\caption{Estimation (E), through the full-shell estimator
Eq.\ref{xifs1}, of the theoretical (absolute value) $\xi(r)$ (T) given
by Eq.\ref{xi-cdmlike}: In this case $R_s=500 > r_{zp}=60$. One may
note that the negative tail is distorted in a non-linear way in order
to satisfy Eq.\ref{xifs}.}
\label{xicdm}
\end{figure}

A similar situation happens when $\cd$ has a power-law behavior inside
a given sample of size $R_s$. (Note that the following argument can be
simply modified to any other functional behavior of $\cd$ in the
regime where $\cd > n_0$). Suppose then that the scale where $\cd
\approx n_0$ is larger than $R_s$ and that
\be
\label{cde} 
\cd = B r^{-\gamma} \ee where $3> \gamma >0$.  Neglecting
fluctuations, the estimation of the sample density from Eq.\ref{ne1}
becomes \be
\label{ne}
n_E= \frac{3B}{3-\gamma} R_s^{-\gamma} 
\ee
so that the estimation of $\xi(r)$ can be written as (again,
neglecting fluctuations)
\be
\label{xine}
\xi_E(r) = \frac{3-\gamma}{3} \left(\frac{r}{R_s}\right)^{-\gamma} -1
\;.  \ee In this case both the scales at which $\xi(r)=1,0$ are
linearly dependent on the sample size $R_s$.

Note that the estimation in Eq.\ref{ne} has been done by assuming that
one can perform a volume average also at the scale of the sample: this
means that one has made an average over different samples of size
$R_s$. In case this is not possible (i.e. the usual situation in
galaxy catalogs) significant deviation from the estimation given by
Eqs.\ref{ne}-\ref{xine} can be found (see \cite{book} for a detailed
explanation of this point). 

It is worth noticing that while statistical noise may change the scale
where $\xi(r_{zp})=0$, it does not change the fact that such a scale
depends on the sample size as long as the conditional density has not
become constant as a function of scale.  However one should note that
for a functional behavior of the type strong power-law correlations
followed by a regime where $\xi(r)$ is very small (or zero or negative
as in the CDM case) the scale $r_{zp}$ can be easily identified by the
scale where a sharp break down from a power law behavior is manifest,
which corresponds to the scale where $\cd
\approx n_0$. This is actually the way in which the constraint imposed
by Eq.\ref{xifs} is evident. The situation where the scale $r_{zp}$
corresponds to a real feature, i.e.  $R_s \gg r_{zp}$, is much more
problematic to be measured and it requires a very careful analysis of
the estimator errors. For example the detection of very small
amplitude correlations can be masked, at least, by Poisson noise going
as $1/\sqrt{N}$.

\section{Comparison with observations}

The characterization of galaxy clustering is usually performed through
the study of the reduced two-point correlation function. The result
found in various galaxy catalogs is that $\xi_E(r) \approx A\times
r^{-\gamma}$ when $\xi_E(r)\gtapprox 1$ with $\gamma=1.7$ and $A$ is a
constant which takes different values in different volume limited
(hereafter VL) subsamples
(e.g. \cite{dp83,davis88,norberg02,zehavi04}).

One should note that a VL is constructed in such a way to contain all
galaxies brighter than a certain absolute magnitude threshold and it
is limited by a distance depending on the apparent magnitude limit of
the galaxy catalog and on the absolute magnitude threshold considered
(e.g., \cite{dp83}). This implies that a VL sample is identified (at
least) by two cuts, one in the distance $R_{VL}$ and one in the
corresponding absolute magnitude $M_{VL}$, the relation between the
two being (at small redshift, neglecting corrections)
\be
\label{vl}
M_{VL}= m_{lim} - 5 \log_{10} R_{VL} -25
\ee 
where $m_{lim}$ is the apparent magnitude limit of the considered
galaxy survey and $R_{VL}$ is measured in Mpc/h. Thus, when one
increases $R_{VL}$ only galaxies with brighter absolute luminosity
(decreasing absolute magnitude $M<M_{VL}$) are included in the
sample. (In latest surveys like SDSS and 2dF, there are two cuts in
apparent magnitude, and thus a VL is identified by two cuts in
absolute magnitude and two in distance: this complicates the
estimation of the depth of the samples but does not introduce a
substantial change in the following discussion).

Given the two parameters $R_{VL}\,,M_{VL}$ defining a VL sample, one
may consider (at least) two different effects which may cause the
amplification of $\xi_E(r)$: (i) a luminosity (or sampling) effect
related to the selection of different class of objects in different VL
samples\footnote{A similar effect happens when the selection is done
on the basis of galaxy color (e.g. \cite{zehavi04}). As there is a
correlation between galaxy color and luminosity, this adds a
complication but no essential change to the logic of our argument.}; 
(ii) a finite-size effect related to the change of the volume of the
samples considered when the absolute magnitude cut is changed. In
other words the variation of the amplitude of $\xi_E(r)$ can be related
to a sampling effect (e.g. to Eq.\ref{approx-xinu}) or to a volume
effect (e.g. to Eq.\ref{xine}): The situation is illustrated in
Figs.\ref{ximodel1}-\ref{ximodel3}.
\begin{figure}
\begin{center}
\includegraphics*[angle=0, width=0.5\textwidth]{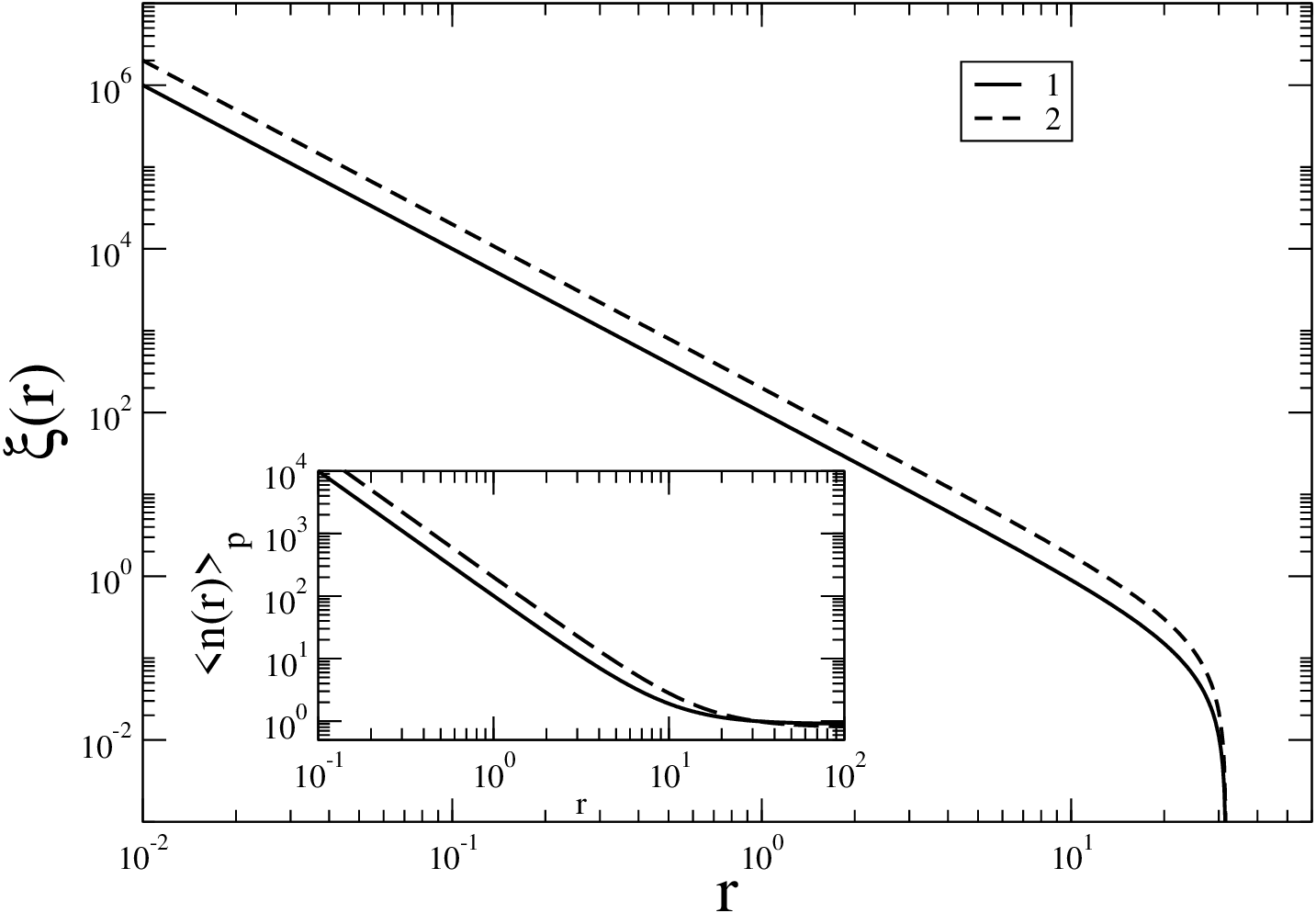}
\end{center}
\caption{Amplification of $\xi_E(r)$ due to a sampling procedure 
similar to what is found in N-body simulation. The underlying 
particle distribution (1) and the selected ones (2) have a 
different amplitude in the regime of strong clustering but
show the same zero-point crossing scale in the reduced correlation
function which has a power-law decay up to a definite scale.
In the insert panel it is shown the corresponding behavior
of the conditional density. Note that the scale $r_{zp}$ coincides 
with the scale where $\cd\approx$ constant.}
\label{ximodel1}
\end{figure}
\begin{figure}
\begin{center}
\includegraphics*[angle=0, width=0.5\textwidth]{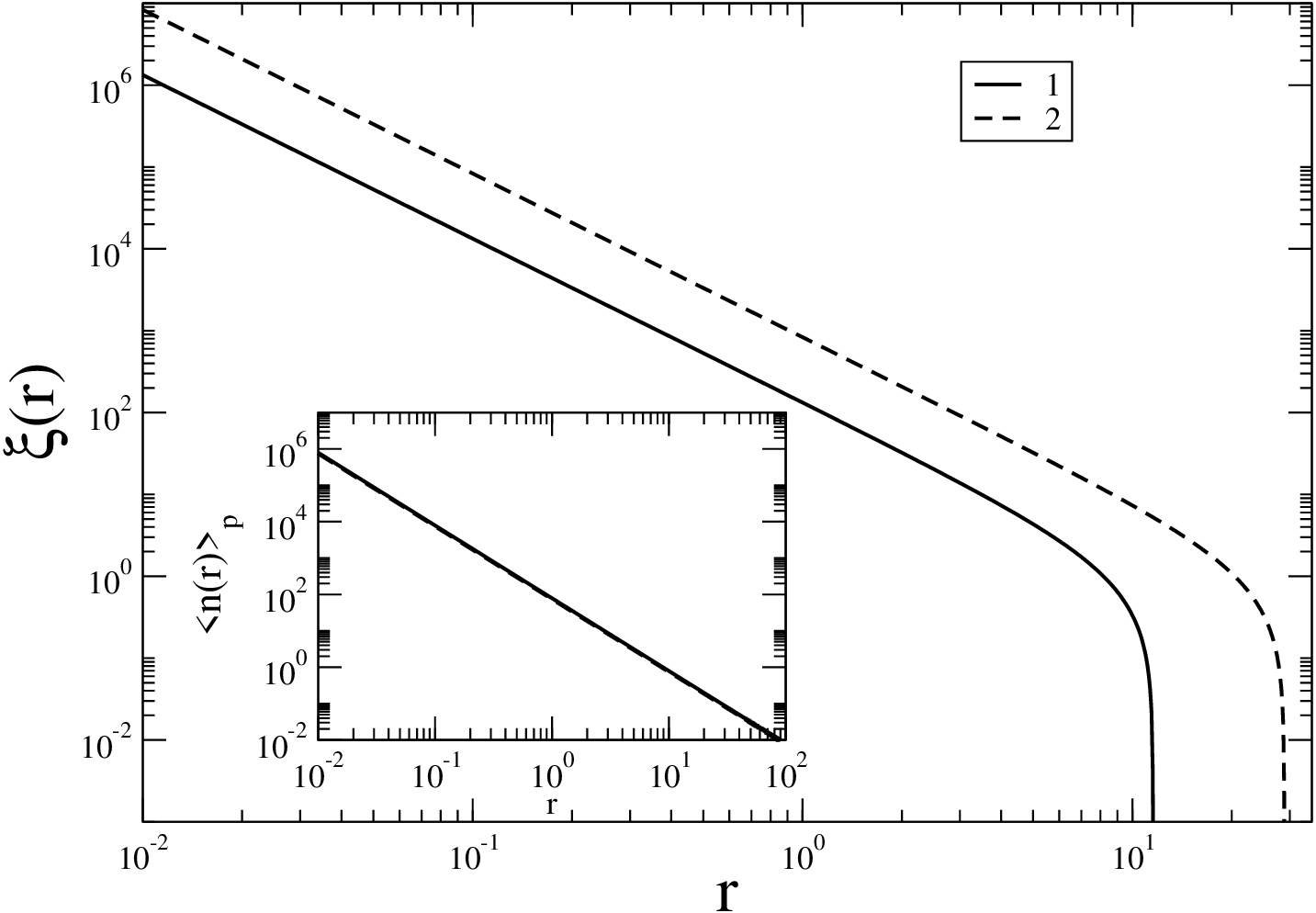}
\end{center}
\caption{In the case the distribution has a conditional density 
with a power-law behavior up to the sample size (1), then the
amplitude of $\xi_E(r)$ and its zero-crossing scale depends linearly on
the sample size (2) if the sample size is larger than the scale where
$\cd$ has a clear flattening toward a constant value. In the insert
panel it is shown the corresponding behavior of the conditional
density: in this case the two lines coincide, the case (2) extending
to larger scales.}
\label{ximodel3}
\end{figure}

Note that the variation of the amplitude of $\xi_E(r)$ (or of its
Fourier conjugate the power spectrum) is usually
(e.g. \cite{dp83,davis88,norberg02,zehavi04}) ascribed to the fact
that galaxies of different luminosity are differently clustered in the
sense that brighter galaxies have a larger amplitude than fainter ones
(this is usually called ``luminosity bias''): i.e.  $A=A(M_{VL})$ is
an increasing function of the absolute luminosity $M_{VL}$ of the
considered galaxies (e.g. \cite{norberg02,zehavi04}). Also for galaxy
clusters a similar variation in the amplitude, although larger, has
been found (e.g. \cite{bs83}) where the variation is ascribed to the
richness of the clusters considered. In brief this variation is
ascribed to some specific ways of sampling the (galaxy or cluster)
point distribution\footnote{Note that (\cite{zehavi04}) made some
specific measurements able to test for finite-size effects, with the
result that large sample fluctuations do alter the amplitude of
$\xi_E(r)$. A discussion of these results can be found in
(\cite{jsetal05}).}. If this would be the case than one should find,
for the zero-crossing length scale $r_{zp}$ a situation analogous to
the one shown in Fig.\ref{ximodel1}: i.e. this scale should be the
same for different objects.

Thus in order to distinguish between the two different mechanics of
amplification of $\xi(r)$ one has an indirect and a direct test.  The
former consists in the study of the stability of the zero-crossing
length scale in different samples, while the latter is represented by
the determination of the conditional density in VL samples. As the
conditional density is not usually estimated (e.g. Zehavi et
al. 2004B, Norberg et al., 2002) I need to consider also the stability
of $r_{zp}$. Below I will comment about the relation with the
measurements of $\cd$ recently performed by Hogg et al. (2005) and the
various determinations summarized in Gabrielli et al. (2004).

It is interesting to briefly review some determinations in redshift
space of the scale $r_{zp}$: in the CfA1 sample $r_{zp}
\approx 20$ Mpc/h (\cite{dp83}); Park et al. (1994) found, in the CfA2
catalog, a larger value of about $r_{zp} \approx 30$ Mpc/h (see their
Fig.10) and Benoist et al.  (1996) found that $r_{zp}$ is not stable
in different VL samples of the SSRS2 survey, changing from 10 to about
50 Mpc/h (see their Fig.1).  More recently it has been found that
$r_{zp} \approx 40$ Mpc/h in the Two degree Field Galaxy Redshift
Survey (\cite{2df}). The latest determination has been performed by
Eisenstein et al. (2005) by considering the Luminous Red Galaxies
sample from the SDSS. This sample covers the largest volume of
universe up to now. In Eisenstein et al. (2005) the zero-point of the
correlation is found to be at a scale of about 120 Mpc/h (see their
Figs.2-3).  Thus it seems that, up to now, in galaxy samples, the
length scale $r_{zp}$ is related to the length scale $r_0$ (defined as
$\xi(r_0)=1$): they are both sample size dependent.  Whether the
latest measurement by Eisenstein et al. (2005) is stable will have to
be shown by the analysis in larger samples.

Note that, as discussed above, one of the main characteristic of the
selection mechanisms usually considered is that the zero-crossing
scale of $\xi(r)$ is invariant under sampling. Thus even if one uses a
very particular kind of objects, results on the zero-crossing scale
have to be the same for any other kind of objects if the difference in
the correlation function (or power spectrum) are explained by a
selection effect similar to what is found in the N-Body
simulations. If the zero-crossing scale is instead not found to be
stable in different samples and thus for different objects, this is a
clear indication that correlation properties are finite-size dependent
in the sense of Fig.\ref{ximodel3}.

The direct test (corresponding to the insert panels in
Figs.\ref{ximodel1}-\ref{ximodel3}) for this has been implicitly
performed by Hogg et al.  (2005) where they measured the (integrated)
conditional density for the same Luminous Red Galaxies sample
considered by Eisenstein et al (2005).  They in fact find that the
conditional density, having a power law behavior with exponent $\gamma
\approx 1$ up to $20\div30$ Mpc/h, shows a slow crossover toward
homogeneity, reaching a constant value at about 70 Mpc/h. These
results support the conclusions drawn here, that the zero-point
crossing scales found in previous and smaller volume surveys is a
finite size effect.  The results by Hogg et al. (2005) are then in
agreement with those of Eisenstein et al. (2005): here we note that
the flattening of $\cd$ occurs at scales comparable to the sample size
and thus this situation requires a careful study of larger samples to
confirm these results over a substantial range of scales (see
discussion in Joyce et al. 2005).

\section{Conclusions} 

The study of the dependence of the zero-crossing scale as a function
of the size of a given sample  is already a vailable test to
distinguish between the different effects producing the variation of
the amplitude of $\xi(r)$.  As long as it is found to be dependent on
the finite sample size, this means that all amplitudes related to
$\xi(r)$ are also finite size dependent. In such a situation a more
clear way to study the problem is represented by the analysis of the
conditional density $\cd$ (see e.g. \cite{book}).  From a review of
the literature it seems that the scale $r_{zp}$ has grown from 20
Mpc/h in the CfA1 sample (\cite{dp83}) to about 120 Mpc/h in the
latest SDSS data (\cite{dje1}). Analogously the scale $r_0$ (defined
as $\xi(r_0)=1$) has grown from about 5 Mpc/h in the CfA1
(\cite{dp83}) to about 13 Mpc/h in the SDSS sample
(\cite{zehavietal2004B}).

This implies that the explanation of the amplitude variation of
$\xi(r)$ by luminosity bias (brighter objects have larger amplitudes)
is untenable. Such a variation can be instead explained as a finite
size effect. To directly test this fact one may simply measure the
conditional density and results for this quantity (Sylos Labini et
al. 1998, Hogg et al., 2005) unambiguously support the fact the the
amplitude variation of $\xi(r)$, or of its zero-crossing length, are
finite-size effects (see Figs.\ref{ximodel1}-\ref{ximodel3}).  This
situation implies that $r_0$ is sample size dependent up to the scale
where $\cd$ has a clear crossover. If one considers such a scale to be
70 Mpc/h, as suggested by Hogg et al. (2005), then $r_0\approx 13$
Mpc/h for galaxies of any luminosity. Note that the prediction of
Eq.\ref{xine} does not apply in this situation as the conditional
density measured by Hogg et al. (2005) shows two different 
behaviors in the strongly clustering regime: a simple power-law up to
about $20$Mpc/h a a slow crossover up to 70 Mpc/h. In this situation
the estimation of $r_0$ has to be done numerically.

The difference between the zero-crossing length scales, found by the
sharp cut-off in a log-log plot of the correlation function, in the
galaxy catalogs extracted from N-body simulations (which is about 30
Mpc/h) and the one detected by (\cite{dje1}) for the largest
observational sample of the SDSS available up to now, is of about a
factor five.  In the situation considered here the zero-point of
$\xi(r)$ is the scale where $\cd
\approx n_0$ and thus this is related to the size of the largest
non-linear structure in the distribution.  This implies that
structures formed in N-body simulations are  smaller than
galaxy structures. This can be directly tested by comparing the scale
where $\cd\approx$ const. in simulations and in galaxy samples (see
discussion in Joyce et al., 2005).

It is important to stress that the conditional density in N-body
simulations (see Fig.\ref{gammasimu}) has a slope of about
$\gamma=-1.7$ while in galaxy catalogs Sylos Labini et al. (1998) and
Hogg et al. (2005) have measured $\gamma=1$. While the analysis of
$\xi(r)$ does not give a clear determination of the slope $\gamma$, as
it is affected by a finite size effect when $\cd$ is a power-law, the
analysis of the conditional density provides with a clear result (see
discussion in Gabrielli et al. 2005). In other words, while the
comparison of $\xi(r)$ in simulations and galaxy samples can be
misleading, this is not the case for $\cd$.

One should also note that in N-body simulations the slope is
determined in real space, while in results in galaxy catalogs
considered here are in redshift space. However the scales involved
(some tens Mpc/h where peculiar velocities are expected to be small)
and the large difference in the slopes found (about 0.7) point toward
a real difference between structures formed in N-body simulations and
observed in galaxy catalogs.

It is worth noticing that the scale $r_{zp}$, in CDM models, is simply
related to the so-called turn-over wave-number of the power spectrum,
i.e. where the power spectrum changes regime from negative to positive
power law.  In this respect, I note that for the determination of the
power-spectrum of density fluctuations a finite size effect in the
amplitude and in the location of the turn-over scale, in a similar way
to what happens for $\xi(r)$, is expected to be present as long as the
distribution has strong clustering inside a given sample
(\cite{sla}). Such a situation allows one to simply relate the results
of Tegmark et al. (2004) for the power spectrum in the SDSS survey, to
the results obtained by the real space correlation function analysis
by Zehavi et al. (2004B).


\begin{acknowledgements}

I am grateful to Michael Joyce, Thierry Baertschiger, Andrea
Gabrielli, Luciano Pietronero, Yuri V. Baryshev and Ravi Sheth for
useful discussions and comments.

\end{acknowledgements}

{}

\end{document}